# SPIN EFFECT ON THE RESONANT TUNNELING CHARACTERISTICS OF A DOUBLE-BARRIER HETEROSTRUCTURES UNDER LONGITUDINAL STRESSES

H. Paredes Gutiérrez[1], S. T. Pérez-Merchancano[2] and G. E. Marques[3]

[1] *Escuela de Física, Universidad Industrial de Santander, A. A. 678, Bucaramanga, Colombia*
[2] *Departamento de Física, Universidad del Cauca, calle 5 4-70, Popayán, Colombia*
[3] *Departamento de Física, Universidade Federal de São Carlos, 13.565-905, São Carlos, São Paulo, Brazil*

Theoretical research on electronic properties in mesoscopic condensed matter systems has focused primarily on the electron charge freedom degrees, while its corresponding spin freedom degrees have not yet received the same attention. Nevertheless nowadays there has been an increment in the number of electron spin-related experiments showing unique possibilities for finding novel mechanisms of information processing and transmission, opening ample fields of opportunities in the theoretical developed of new models. In this spirit we have calculated the resonant tunneling characteristics curves in double-barrier heterostructures of GaAs-Ga$_{1-x}$Al$_x$As under external stress and considering two charges with spin half. The resonant tunneling study has been carried out by means of the diagrammatic techniques for non equilibrium processes following the model proposed by Keldysh also a simple one-band tight-binding Hamiltonian is adopted in the theoretical framework. We have compared our results of the spin-tunneling with previous ones reported in literature.

**Key words**: Transport; Resonant tunneling; Low dimensionally system.

## Introduction

The effect of symmetry on the energy band structure of crystals with zinc blende structure can be readily derived using different methods of the solid state physics and of quantum mechanics. Recent extensive studies of the semiconductors properties of (Ga,Al)As, which has the zinc blende structure and preliminary cyclotron resonance investigations, have indicated a need for a more thorough understanding of the possible energy band structure of a zinc blende type crystal. However, the physics of spin-dependent tunneling phenomena has attracted a rapidly growing interest due to its potential application in spintronics [1], that permit the study and design of spin manipulated devices used as information processors, quantum computing elements, spin polarized diodes, spin-valve read heads and electro-optical modulators to name just a few. Glazov, et al [2], present a theory of spin dependent tunneling through a symmetrical double-barrier heterostructure (DBH) based on zinc blende lattice semiconductor compounds. The Dresselhaus spin-orbit interaction couples spin states and space motion of conduction electrons that leads to spin splitting of the resonant level depending on the in-plane electron wave vector. Recently, Rui Wang, et al [3], studied quantum transport through a quantum dot array coupled with two semi-infinite leads using the tight-binding Hamiltonian which is spin-dependent due to external magnetic field. They have calculated both reflection and transmission probabilities using the transfer matrix method which depends on the spin-polarization. On the other hand other authors have proposed that the asymmetric nonmagnetic semiconductor barrier itself could serve as a spin filter. It was demonstrated that the spin-dependent electron reflection in equivalent interfaces resulted in the dependence of the tunneling transmission probability on the orientation of electron spin. This effect is caused by interface-induced Rashba spin orbit coupling and can be substantial for resonant tunneling through asymmetric double-barrier heterostructures.

In this work we have calculated the resonant tunneling characteristics curves in double-barrier heterostructures of GaAs-Ga$_{1-x}$Al$_x$As under external stress and considering two charges with half spin. The resonant tunneling study has been carried out by means of the diagrammatic technique for no equilibrium processes following the model proposed by Keldysh [4] also a simple one-band tight-binding Hamiltonian is adopted in the theoretical framework. We have compared our results of the tunneling-via-spin with previous ones reported in literature [2].

**Theoretical Model**

Consider a simple one-band tight-binding model for the spin dependent resonant tunneling (neglecting of spin-orbit interaction) and under external stress through a DBH of GaAs-$Ga_{1-x}Al_xAs$. The Hamiltonian of the system has the form

$$H = \sum_{i,\sigma} \varepsilon_i c_{i\sigma}^+ c_{i\sigma} + \sum_{i,j,\sigma} V_{ij} c_{i\sigma}^+ c_{j\sigma} + \sum_{\sigma,\sigma'} A_{\sigma,\sigma'} c_\sigma^+ c_{\sigma'}, \quad (1)$$

where $\varepsilon_i$ is the diagonal energy, $c_{i,\sigma}^+$ ($c_{i,\sigma}$) is the creation (annihilation) operator of an electron with spin spin-up ($S^+$) and spin-down ($S^-$) in either the well or barrier, $V_{ij}$ is the hopping neighboring sites $i$ and $j$. In Eq. (1)

$$A_{\sigma,\sigma'} = \frac{e\hbar}{2m^*} \vec{\sigma} \cdot \vec{B}, \quad (2)$$

where $e$ the electron charge, $m^*$ is the electronic effective mass, $\vec{\sigma}$ represents the Pauli matrices and $\vec{B}$ is the magnetic field applied parallel to the current direction.

In order to treat the Green functions (GF) in this nonequilibrium situation we make use of the Keldysh scheme, the full system is decoupled into two equilibrium ones [right (R) and left (L)] and associated GF are obtained. Renormalized dressed GF for the nonequilibrium case can be derived, via Dyson equation, by linking the two subsystems with an perturbed Hamiltonian. The average current, at $T = 0K$, induced in the system is given by [5,6]

$$\langle I \rangle = \frac{2eT}{\hbar} \int_{-\infty}^{+\infty} d\omega [G_{01}^{+-}(\omega) - G_{10}^{-+}(\omega)], \quad (3)$$

which can be written in terms of the density of states of the two subsystems in equilibrium, $\rho_{RL}(\hbar\omega)$, as

$$I = \frac{4\pi^2 eT^2}{\hbar} \int_{\mu_R}^{\mu_L} \frac{\rho_L(\hbar\omega)\rho_R(\hbar\omega)d(\hbar\omega)}{|\Lambda(\hbar\omega)|^2}, \quad (4)$$

where $T = V_{01} = V_{10}$ and

$$|\Lambda|^2 = (1 - g_{LL}^a g_{RR}^a V^2)(1 - g_{LL}^r g_{RR}^r V^2), \quad (5)$$

with $g_{LL(RR)}^{a(R)}$ corresponding to the advanced (retarded) GF of the left and right subsystem. We have considered $\mu_R < \mu_L$, $\mu_L$ and $\mu_R$ as being the chemical potentials of the injector and collector located on the left and right-hand sides of the system, respectively. We refer to the left chemical potential uniquely as the Fermi energy.

**Results**

We use the GaAs electronic effective mass $m^* = 0.067 m_o$, $m_o$ being free electron mass, a lattice constant $a = 2.82 \text{Å}$, and a Fermi level of $75 meV$, which corresponds approximately to emitter doping of the order $1.0 \times 10^{18} cm^{-3}$. Here we consider a barrier height ($\Delta E$) of $249 meV$ and barrier and well widths $42 \text{Å}$. The current versus voltage characteristic (I-V) of a GaAs-Ga$_{1-x}$Al$_x$As DBH for two different longitudinal stress $S_{HP} = 0 kbar$ and $S_{HP} = 30 kbar$ with spin-up ($S^+$) and spin-down ($S^-$), is presented in Fig 1(a). We can observe that the presence of the stress affects the spin properties of the resonant tunneling peak. Note the shift at higher (lower) voltage for $S^-$ ($S^+$) with increasing stress. In this figure also it is observed that the intensity of current increases quickly and that the potential difference may reach a value of $0.03V$. Later, the current reaches the negative differential resistance region of the double barrier. About 0.2 V, the increase in current is observed again because new resonant states appear in the system. This situation occurs with and without hydrostatic pressure in the model used here. Similarly, we can notice important changes in the behavior of the current of the system when the hydrostatic pressure changes from zero to 30 Kb, showing that with smaller potential we can obtain an increment in the spin up or spin down current peaks when the hydrostatic pressure increases.

To describe the spin polarization of the current and its dependence on the system parameters, we defined the current polarization as, $P = I_{Down}/(I_{up} + I_{Down})$. In fig 1(b), we present the polarization $P$ versus voltage $V$, where the structural parameters are the same as in Fig 1(a). We can see that, as the voltage increases, the number of electrons with $S^-$ transmitted increase until a voltage of $0.03V$, where 96% of the concentration of electrons go through the barriers causing an effective voltage control of spin polarized current. Also, we observe that when the voltage increases, an inversion in the spin polarized current occurs, changing the number of electrons tunneling through the barriers and the device may behaves as an spin filter. When the voltage reaches $0.18V$, we see an increase of the current and the amount of electrons transmitted with spin-up polarization. The polarization curve is shifted to lower voltages as the stress increases.

In fig 2, we plot of the transmission probability as a function of the energy for the spin-down (stress, $S_{HP} = 0 kbar$, dashed line) and spin-up (stress 30 kbar., solid line) electrons, for the same structure as in Fig 1. We can observe that the presence of the stress and spin causes a shift in the transmission peaks, to low energy. This is due to the presence of the spin-field interaction, the third term in the Hamiltonian (cf. Eq. (1)), in qualitative agreement with theoretical results [2]. The peaks, at about 0.09eV ($S_{HP}^+ = 0 kbar$), and 0.10eV ($S_{HP}^- = 30 kbar$), are resonant energies below the tip of the barrier at 0.249eV, and negative conductances ocurr around 0.18V ($S_{HP}^+ = 0 kbar$) and 0.20V ($S_{HP}^- = 30 kbar$) [see fig 1 (a)] and is more significant in the thicker barriers. This applied voltage corresponds to the first transmission probability peak approaching the Fermi energy in the trapezoidal well.

**Conclusion**

In summary, we have studied spin-up and spin-down of resonant tunneling in DBH in the presence external stress via the I-V characteristics of the system. We have followed a theoretical description based on Keldysh's nonequilibrium Green function, which is adequate to describe transport properties. The current under the presence of stress and the spin has effects on the resonant tunneling peaks, that show a shifts to higher and lower voltages. The transmission probability for electrons tunneling through this kind of system is obtained, which can be used to discuss the spin-dependent transport of electron in double barrier heterostructure. Our results are in good agreement with other theoretical results.

**Acknowledgements**

The authors of this work thankful the partial support a the Industrial University of Santander (UIS) through the "Dirección General de Investigaciones (DIF Ciencias, Cod. 5124)", the Excellence Center of Novel Materials ECNM, under Contract No. 043-2005 and the Cod. No. 1102-05-16923 subscribed with COLCIENCIAS, FACNED and VRI project ID 2193 Universidad del Cauca. We express our thanks to Dr. I. D. Mikhailov for useful discussions about resonant tunneling.

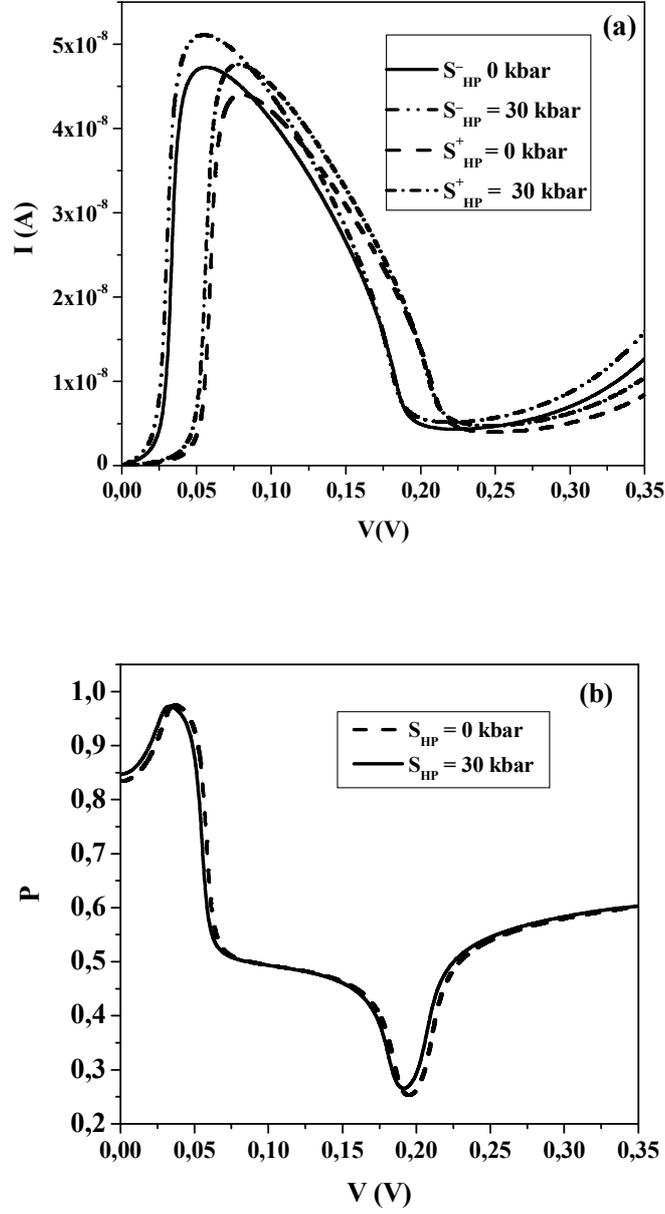

Figure 1 (a) Current vs. voltage for DBH under two different longitudinal stress for spin-up and spin-down carriers. The height of the barrier potential is $\Delta E = 249 meV$ and the well and the barriers have 42 Å, respectively. (b) The spin polarization vs. bias voltage for DBH for the same structure as in Fig 1 (a).

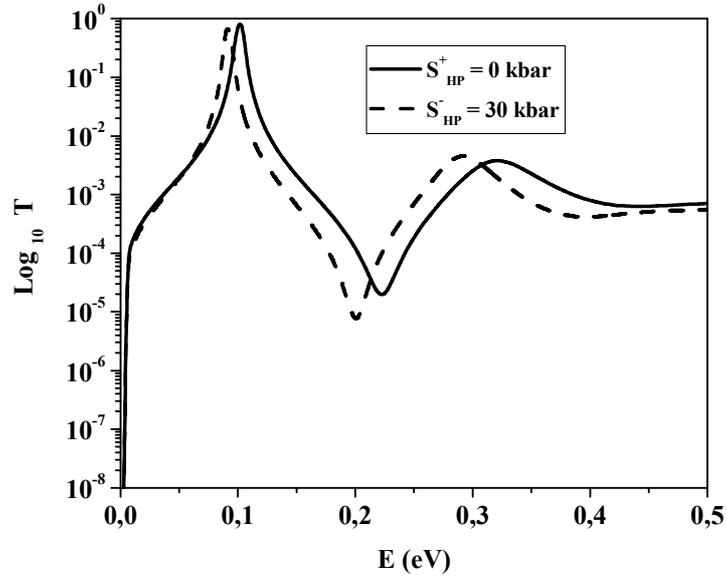

Figure 2. Logarithm of the transmission probability $T$ versus energy $E$ for DBH under two different longitudinal stress for spin-up and spin-down carriers. The height of the barrier potential is $\Delta E = 249 meV$ and the well and the barriers have $42$ Å, respectively.